\documentclass[prd,twocolumn,twoside,preprintnumbers,superscriptaddress,nofootinbib,natbib,bm,amsrefs]{revtex4-1}
\usepackage{graphicx}
\usepackage{hyperref}
\usepackage{url}

\usepackage{xcolor}
\usepackage{amsmath, amstext, amsfonts,amssymb, amsthm}
\usepackage{comment}
\usepackage[utf8]{inputenc}
\usepackage{multirow}
\newcommand{\ov}{\overline}

\newcommand{\GeV}{{\rm GeV}}

\definecolor{BlueViolet}{rgb}{0.2, 0.00, 0.7}
\definecolor{Blue}{rgb}{0.15, 0.00, 0.9}
\definecolor{lightblue}{rgb}{0.15, 0.35, 0.95}
\definecolor{kitgreen}{rgb}{0,
0.58823 
, 0.50980 
}

\newcommand{\Eprint}[1]{\href{#1}}

\definecolor{lb}{rgb}{.74,.83,.9}
\definecolor{ly}{rgb}{1,.92,.8}
\definecolor{lr}{rgb}{.98,.85,.87}

\begin{document}
\preprint{PSI-PR-23-41, ZU-TH 73/23, P3H-23-078, TTP23-051}
\title{Accumulating Hints for Flavour Violating Higgses at the Electroweak Scale}

\author{Andreas Crivellin}
\email{andreas.crivellin@psi.ch}
\affiliation{Physik-Institut, Universität Zürich, Winterthurerstrasse 190, CH–8057 Zürich, Switzerland}
\affiliation{Paul Scherrer Institut, CH–5232 Villigen PSI, Switzerland}

\author{Syuhei Iguro}
\email{igurosyuhei@gmail.com}
\affiliation{Institute for Theoretical Particle Physics (TTP), Karlsruhe Institute of Technology (KIT), Wolfgang-Gaede-Str.\,1, 76131 Karlsruhe, Germany}
\affiliation{Institute for Astroparticle Physics (IAP), KIT, Hermann-von-Helmholtz-Platz 1,\\ 76344 Eggenstein-Leopoldshafen, Germany
}
\affiliation{Institute for Advanced Research (IAR), Nagoya University, Nagoya 464--8601, Japan}

\begin{abstract}
We show that supplementing the Standard Model by only a second Higgs doublet, a combined explanation of $h\to e\tau$, $h\to \mu\tau$, $b\to s \ell^+ \ell^-$, the $W$ mass and $R({D^{(*)}})$ as well as the excess in $t\to bH^+(130\,{\rm GeV})\to b\ov{b}c$ is possible. While this requires flavour violating couplings, the stringent bounds from e.g.~$\mu\to e\gamma$, $\tau\to \mu\gamma$, $B_s-\bar B_s$ mixing, $b\to s\gamma$, low mass di-jet and $pp\to H^+H^-\to \tau^+\tau^-\nu\bar\nu$ searches can be avoided. However, the model is very constrained, it inevitably predicts a shift in the SM Higgs coupling strength to tau leptons as well as a non-zero $t\to hc$ rate, as indeed preferred by recent measurements. We study three benchmark points providing such a simultaneous explanation and calculate their predictions, including collider signatures which can be tested with upcoming LHC run-3 data.
\end{abstract}
\maketitle

\section{Introduction}
\label{sec:intro}

The Standard Model (SM) describes the known fundamental constituents of matter and their interactions at sub-atomic scales. It has been extensively tested and verified by a plethora of measurements~\cite{ParticleDataGroup:2022pth} and the discovery of the Brout-Englert-Higgs boson~\cite{Higgs:1964ia,Englert:1964et,Higgs:1964pj,Guralnik:1964eu} at the LHC~\cite{ATLAS:2012yve,CMS:2012qbp}, which has, in fact, properties~\cite{Langford:2021osp,ATLAS:2021vrm,CMS:2022dwd,ATLAS:2022vkf} in agreement with the SM expectations, provided its last missing puzzle piece. 
However, these results do not exclude the existence of additional scalars, if the SM-Higgs signal strengths are not significantly altered (i.e.~the mixing with the new scalars is sufficiently small) and their contribution to the $\rho$ parameter ($\rho=m_Z^2\cos^2\theta_W/m_W^2$) does not violate the experimental bounds. In fact, several indirect and direct hints suggest the existence of new Higgses (see Ref.~\cite{Crivellin:2023zui} for a recent review). 

In this article, we consider the simple and motivated option of extending the SM by a single Higgs doublet, i.e.~a two-Higgs-doublet model (2HDM), see Ref.~\cite{Branco:2011iw} for a review. We will focus on flavour-violating signatures motivated by an interesting set of anomalies, i.e.~deviations from the SM predictions: non-zero rates of $t\to bH^+(130)\to b\ov{b}c$, $h\to e\tau$ and $h\to \mu\tau$ as well as the deviations from the SM predictions in $b\to s \ell^+ \ell^-$, the $W$ mass and $R({D^{(*)}})$. For an explanation, flavour violation is clearly required (except for the $W$ mass), and we will thus consider the 2HDM with generic Yukawa couplings~\cite{Hou:1991un,Chang:1993kw,Liu:1987ng,Cheng:1987rs,Savage:1991qh,Antaramian:1992ya,Hall:1993ca,Luke:1993cy,Atwood:1995ud,Atwood:1996vj,Botella:2015hoa,Herrero-Garcia:2016uab} as a minimal model with the potential of explaining these measurements. However, there are various bounds from flavour and collider observables which must be respected, such that the model is very constrained and it is {\it{a priori}} not clear if a combined explanation is possible.

The layout of this article is given as follows: In Sec.~\ref{sec:model} we introduce our model. Then in Sec.~\ref{sec:explanation} we consider the status of the anomalies and the relevant bounds together with our NP contributions. 
In Sec.~\ref{sec:pheno} we propose the benchmark points that can resolve the anomalies and calculate their predictions before we conclude in Sec.~\ref{sec:conclusion}.

\section{G2HDM}
\label{sec:model}

In the 2HDM with generic Yukawa couplings (G2HDM), also called the type-III 2HDM, one can work in the so-called Higgs basis where only one Higgs doublet acquires a nonzero vacuum expectation value (VEV)~\cite{Davidson:2005cw} such that 
\begin{eqnarray}
  H_1 =\left(
  \begin{array}{c}
    G^+\\
    \frac{v+\phi_1+iG^0}{\sqrt{2}}
  \end{array}
  \right),~~~
  H_2=\left(
  \begin{array}{c}
    H^+\\
    \frac{\phi_2+iA}{\sqrt{2}}
  \end{array}
  \right).
\label{HiggsBasis}
\end{eqnarray}
Here, $G^+$ and $G^0$ are would-be Goldstone bosons, and $H^+$ and $A$ are the charged Higgs and the CP-odd Higgs boson, respectively, with $v\approx 246\,$GeV. The Yukawa couplings can then be written as
\begin{align}
\mathcal{L}_Y = & - \ov Q_L^i\!\left( {{H_1}y_d^i + {H_2}\rho _d^{ij}} \right)\!d_R^i - \ov L_L^i\!\left( {{H_1}y_\ell^i + {H_2}\rho _\ell ^{ij}} \right)\!e_R^i\nonumber\\
& - \ov Q_L^i{({V^\dag })^{ij}}\left( {{{\tilde H}_1}y_u^j + {{\tilde H}_2}\rho _u^{jk}} \right)u_R^j,
\end{align}
where $i$, $j$, and $k$ are flavour indices, and $\tilde H_{1,2}=i \tau_2 H^*_{1,2}$ with $\tau_2$ being the second Pauli matrix. We now perform the rotation
\begin{align}
\begin{pmatrix}\phi_1 \\ \phi_2 \end{pmatrix} = \begin{pmatrix} \cos \theta_{\beta \alpha} & \sin \theta_{\beta \alpha}   \\ -\sin \theta_{\beta \alpha} & \cos \theta_{\beta \alpha} \end{pmatrix}
\begin{pmatrix}H \\ h \end{pmatrix}, 
\end{align}
to go to the mass eigenstates $h$ and $H$ for the neutral Higgses, where $h$ is SM-like. Furthermore, writing $Q=(V^\dagger u_L,d_L)^T$, where $V$ is the Cabbibo-Kobayashi-Maskawa (CKM) matrix~\cite{Cabibbo:1963yz,Kobayashi:1973fv}, we arrive at the mass eigenbasis for the fermions with $y_i^f=\sqrt{2}m_{f_i}/v$ ($m_{f_i}$ denoting the fermion masses).

Note that $\rho_f^{ij}$ is independent of the fermion masses, i.e.~contains 9 complex parameters each for $f=u,d,\ell$. The off-diagonal elements of $\rho_d$ are stringently constrained by meson mixing and decays and we will thus disregard them. We will rather consider the minimal scenario where $\rho_u^{tt}$, $\rho_u^{tc}$, $\rho_\ell^{\tau\tau}$, $\rho_\ell^{\mu\tau}$, and $\rho_\ell^{e\tau}$ are the only nonzero entries. In addition, we consider $m_{H^\pm}$, $m_H$, $m_A$ and Higgs mixing parameter $c_{\beta\alpha}\equiv \cos \theta_{\beta \alpha}$ as free parameters (with relevant impact on the phenomenology) while we disregard CP-violation in the Higgs potential.

\section{Observables}
\label{sec:explanation}

Let us now discuss the different anomalies, the corresponding contributions in the G2HDM as well as the constraints from various other observables.

\subsection{$t\to bH^+(130)\to b\ov{b}c$ }

The ATLAS run-2 analysis~\cite{ATLAS:2023bzb} reported an excess in $t\to b H^+\to b\ov{b}c$ with a global (local) significance of $2.5\,(3.0)\sigma$ at $m_{H^+}\approx 130\,\GeV$ with Br$(t\to bH^+)\times{\rm Br}(H^+\to   \ov b c)=(0.16\pm0.06)\%$.\footnote{The analogous CMS result is available only with run-1 data~\cite{CMS:2018dzl} and hence the sensitivity is not competitive. See Refs.~\cite{Akeroyd:2022ouy,Bernal:2023aai} for alternative explanations of the $m_{H^+}\approx 130\,\GeV$ excess.}
In our model we have 
\begin{align}
    {\rm{Br}}(t\to bH^+)=\frac{m_t|\rho_u^{tt}|^2}{16\pi\Gamma_t}\!\left(1-\frac{m_{H^+}^2}{m_t^2}\right)^{\!2}
    \!\!\approx 0.16 \left( \frac{|\rho_u^{tt}|}{0.06} \right)^{\!2}{\!\%}\,,
\end{align}
where we set $m_{H^+}=130\,\GeV$ on the right-handed side of the equation. As we will see later, the numerically relevant couplings for the decay of $H^+$ are $\rho_u^{tc}$ and $\rho_\ell^{\ell\tau}$, with $\ell=e,\,\mu,\,\tau$, such that
\begin{equation}
    {\rm Br}(H^+\to c \ov{b})\approx \frac{3|\rho_u^{tc}|^2}{3|\rho_u^{tc}|^2+ \sum \limits_{\ell^\prime}|\rho_\ell^{\ell^\prime\tau}|^2}\,.
\end{equation}  

\subsection{$h\to e\tau,\mu\tau$}

While the previous run-1 excess in $h\to\mu\tau$~\cite{CMS:2015qee,ATLAS:2015cji} was not confirmed by run-2 data, the latest ATLAS and CMS results show again indications for nonzero $h\to e\tau$ and $h\to \mu\tau$ rates~\cite{CMS:2021rsq,ATLAS:2023mvd}. The combined significance is $2.4\sigma$ ($\Delta \chi^2=8.3$ with dof=2) with the best fit values Br$(h\to e\tau)\simeq {\rm{Br}}(h\to \mu\tau)\approx0.08\%$ while 2.3$\sigma$ for $h\to e\tau$ and 1.3$\sigma$ for $h\to \mu\tau$ are obtained for the separated measurements.

We have at tree-level 
\begin{align}
    {\rm{Br}}(h\to l\tau)&=\frac{c_{\beta\alpha}^2 m_h }{16\pi^2\Gamma_h}\left(|\rho_\ell^{l\tau}|^2+|\rho_\ell^{\tau l}|^2\right)\nonumber\\
    &\approx 0.06 \left(\frac{c_{\beta\alpha}\sqrt{|\rho_\ell^{l\tau}|^2+|\rho_\ell^{\tau l}|^2} }{10^{-3}}\right)^2\%,
    \label{eq:t_bbc}
\end{align}
where $l=e,\,\mu$.

\subsection{$b\to s \ell^+ \ell^-$}

Recent global $b\to s \ell^+ \ell^-$ fits favour $C_9^U\approx -1$ at the $5\sigma$ level~\cite{Buras:2022qip,Neshatpour:2022pvg,Gubernari:2022hxn,Ciuchini:2022wbq,Alguero:2023jeh,Wen:2023pfq,Capdevila:2023yhq}\footnote{The main drivers for this preference for NP are $P_5^\prime$~\cite{Descotes-Genon:2012isb,LHCb:2015svh,LHCb:2020lmf,LHCb:2020gog}, the total branching ratio and angular observables in $B_s\to\phi\mu^+\mu^-$~\cite{LHCb:2015wdu,LHCb:2021zwz,LHCb:2021xxq} as well as the Br$(B\to K\mu^+\mu^-)$~\cite{LHCb:2014cxe,LHCb:2016ykl,Parrott:2022zte}, which are fully compatible with semi-inclusive observables~\cite{Isidori:2023unk}.}. This means that lepton flavour universal NP with vectorial couplings to lepton and left-handed couplings to bottom and strange is required.

In our model the charm loop contributes to $C_9^U$ via an off-shell photon penguin~\cite{Jager:2017gal,Bobeth:2014rda,Iguro:2018qzf,Crivellin:2019dun,Kumar:2022rcf,Iguro:2023jju} and we obtain~\cite{Crivellin:2019dun},
\begin{align}
    \Delta C_9^U(\mu_b)\approx &-0.52\left(\frac{|\rho_u^{tc}|^2-|\rho_u^{cc}|^2}{0.5^2}\right)+0.50\left(\frac{\rho_u^{tc*}\rho_u^{cc}}{0.01}\right).
    \label{eq:C9U}
\end{align}
We see that a sizable coupling $\rho_u^{tc}$ is necessary if $\rho_u^{cc}\approx0$ is assumed while the product $\rho_u^{tc*}\rho_u^{cc}$ has a CKM enhancement w.r.t.~the SM.

\subsection{$R({D^{(*)}})$}

The long-standing $3\sigma$--$4\sigma$ discrepancy in $B\to D^{(*)} \tau\nu$~\cite{HFLAV:2022pwe} can be solved by a charged Higgs contribution~\cite{Crivellin:2012ye,Crivellin:2013wna,Cline:2015lqp,Crivellin:2015hha,Lee:2017kbi,Iguro:2017ysu,Martinez:2018ynq,Fraser:2018aqj,Athron:2021auq,Iguro:2022uzz,Blanke:2022pjy,Ezzat:2022gpk,Fedele:2022iib,Das:2023gfz} if $m_{H^\pm}\lesssim 400\,\GeV$ avoiding the constraints from $\tau\nu$ searches~\cite{Iguro:2018fni,Iguro:2022uzz}.\footnote{Note that the bound from Br($B_c\to\tau\nu$) is relaxed once the charm mass uncertainty and considerable $p_T$ dependence of fragmentation function of $b\to B_c$ is taken into account~\cite{Alonso:2016oyd,Celis:2016azn,Blanke:2018yud,Aebischer:2021ilm} such that our model can explain the central value of $R(D^{(*)})$. While in principle, the scalar operator changes the differential distributions in $B\to D^{(*)}\tau\nu$~\cite{Sakaki:2014sea,Celis:2016azn,Iguro:2017ysu}, we will not consider these constraints since the theory prediction significantly depends on the form-factors used. In fact, recent lattice results from HPQCD~\cite{Harrison:2023dzh} and Fermi-MILC~\cite{FermilabLattice:2021cdg} have a mild tension with Belle (II) data~\cite{Fedele:2023ewe,Penalva:2023snz} while JLQCD agrees with the measurement~\cite{Aoki:2023qpa,Martinelli:2023fwm}. Furthermore, correlations among the bins of the differential distributions are not provided in both Belle and BaBar papers.}
Our NP contribution is given by
\begin{align}
    C_{S_L}^{\ell^\prime \ell}(\mu_b)
    =& F_{RG}\left( \frac{\rho_u^{tc*}\rho_\ell^{\ell\ell^\prime*}}{m_{H^+}^2} \right)\bigl{/} \left(\frac{4G_F V_{cb}}{\sqrt{2}}\right)\nonumber\\
    \approx& 0.67\left(\frac{\rho_u^{tc*}\rho_\ell^{\ell\ell^\prime*}}{0.01}\right)\left(\frac{130\,\GeV}{m_{H^+}}\right)^2,
    \label{Eq:RD_simplified}
\end{align}
where $F_{RG}\approx1.5$ accounts for the renormalization running effect (RGE)~\cite{Alonso:2013hga,Jenkins:2013wua,Gonzalez-Alonso:2017iyc,Aebischer:2017gaw}. For the numerical analysis we use Ref.~\cite{Blanke:2018yud} to calculate $R(D^{(*)})$, which is consistent with the recent update of Ref.~\cite{Iguro:2022yzr,Iguro:2020cpg}, and use the HFLAV SM prediction~\cite{HFLAV:2022pwe}. Note that in order to explain $R({D^{(*)}})$ at the 1$\sigma$ level and non-interfering effect, either via an imaginary part of $C_{S_L}^{\tau \tau}$ and/or $C_{S_L}^{\tau l}$ is needed.

\subsection{Charged-lepton flavour violation}

The product $\rho_\ell^{e\tau}\rho_\ell^{\mu\tau}$ induces potentially dangerous $\mu\to e$ transitions at the one-loop level. Furthermore, if in addition $\rho_\ell^{\tau e}$ and $\rho_\ell^{\tau \mu}$ were nonzero large $\tau$-mass enhanced contributions to the magnetic operator would arise. However, even if  $\rho_\ell^{\tau l}$ is set to zero we have for $c_{\beta\alpha}=0.1$, $m_\phi=200\,\GeV$ and $m_{H^+}=130\,\GeV$~\cite{Omura:2015xcg}\footnote{We checked that the $\rho_u^{tt}$ induced two-loop Barr-Zee contribution is negligible in our scenario.} 
\begin{align}
    {\rm Br}(\mu\to e\gamma)\approx2.8\times10^{-13} \left(\frac{|\rho_\ell^{e\tau}\rho_\ell^{\mu\tau*}|}{7\times 10^{-5}}\right)^2,
\end{align}
where $\phi=H,\,A$ which can be compared with the current upper limit of ${\rm Br}(\mu\to e\gamma)\le 3.1\times 10^{-13}$ at $90\,\%$~\cite{Cattaneo:2023iua}. Note that the $c_{\beta\alpha}$ dependence is mild for $c_{\beta\alpha}\ll 1$. Moreover, a correlation among $\mu \to e \gamma$, $\mu \to 3 e$ and $\mu\to e$ conversion~\cite{Crivellin:2014cta,Calibbi:2017uvl} can be found
\begin{align}
    {\rm{Br}}(\mu\to e\gamma)\approx 140\!\times \!{\rm{Br}}(\mu\to 3e)\approx 420\!\times \!{\rm{Cr}}(\mu \rm{Al}\to e \rm{Al}),
\end{align}
where Cr corresponds to the conversion rate for which improved measurements are foreseen in the near future~\cite{Blondel:2013ia,Baldini:2013ke,Mu2e:2014fns,COMET:2018auw}. Similarly $\tau\to l \gamma$ is induced by the product $\rho_\ell^{\tau\tau} \rho_\ell^{l\tau}$, but the predicted Br is at least one order smaller than the projected Belle II sensitivity~\cite{Belle-II:2018jsg}.

\subsection{$b\to s \gamma$ and $B_s-\bar B_s$ mixing}

$b\to s\gamma$ and $B_s-\bar B_s$ mixing give relevant constraints on $\rho_u^{tc}$ and $\rho_u^{tt}$. Adopting the global fit of Ref.~\cite{Wen:2023pfq} we find $-0.035\lesssim{\rm{Re}} \Delta C_7(\mu_b)\lesssim 0.037$ at the $2\sigma$ level.\footnote{Reference~\cite{Alguero:2021anc} finds allowed values down to $-0.04$ at $2\sigma$ level.} We obtain the semi-analytic formula for one-loop charged Higgs contribution \cite{Crivellin:2019dun}
\begin{align}
   \Delta C_7(\mu_b)\approx-0.03\left(\frac{|\rho_u^{tc}|}{0.5}\right)^2-0.008\left(\frac{|\rho_u^{tt}|}{0.5}\right)^2
   \label{eq:C7},
\end{align}
meaning that the G2HDM interferes constructively with the SM.

Using the input of Ref.~\cite{Crivellin:2023saq} we obtain the allowed range of $-0.09\le R_{B_s}\equiv\Delta M_{B_s}^{\rm{G2HDM}}/\Delta M_{B_s}^{\rm{SM}}\le0.07$.
This has to be compared to~the charged Higgs contribution which is lengthy and hence omitted \cite{Iguro:2017ysu}. 
Similarly, we can consider $\Delta M_{B_d}$ and $\Delta \Gamma_{B_{s(d)}}$ (see, e.g.~the online update of Ref.~\cite{UTfit:2007eik}), however, the constraints are less stringent and also $D^0-\bar D^0$ and kaon mixing are not relevant for our minimal coupling structure~\cite{Iguro:2019zlc}.

\subsection{$t\to c h(A,H)$}

For the decay of a top to a charm quark and SM Higgs, we have
\begin{align}
    {\rm Br}(t\to hc)&=\frac{ m_tc_{\beta\alpha}^2(|\rho_u^{tc}|^2+|\rho_u^{ct}|^2)}{64\pi \Gamma_t}\left(1-\frac{m_h^2}{m_t^2}\right)^2\nonumber\\
    &\approx 2.4\times 10^{-4}\left(\frac{\rho_u^{tc} c_{\beta\alpha}}{0.05}\right)^2.
\end{align}
Similarly, the rates for $A$ and $H$ are obtained by replacing $m_h$ and $c_{\beta\alpha}$ with $s_{\beta\alpha}$ and $m_{H,A}$, respectively. The current ATLAS upper limit is set as ${\rm Br}(t\to hc)\le4.0\times 10^{-4}$ at 95$\%\;$CL~\cite{ATLAS:2023ujo}. Note that this is weaker than the expected limit of $2.4\times 10^{-4}$ at $95\%$ such that a non-zero rate is in fact preferred. Moreover CMS finds ${\rm Br}(t\to hc)\le3.7\times 10^{-4}$ at $95\%\;$CL, compared to an expected sensitivity of $ 3.5\times 10^{-4}$~\cite{CMS:2023ufv}. The high-luminosity LHC (HL-LHC) can probe Br$(t\to hc)\le1.1\times 10^{-4}$~\cite{TheATLAScollaboration:2013nbo,ATLAS:2016qxw}.

\begin{figure*}[t]
\begin{center}
 \includegraphics[width=0.45\textwidth]{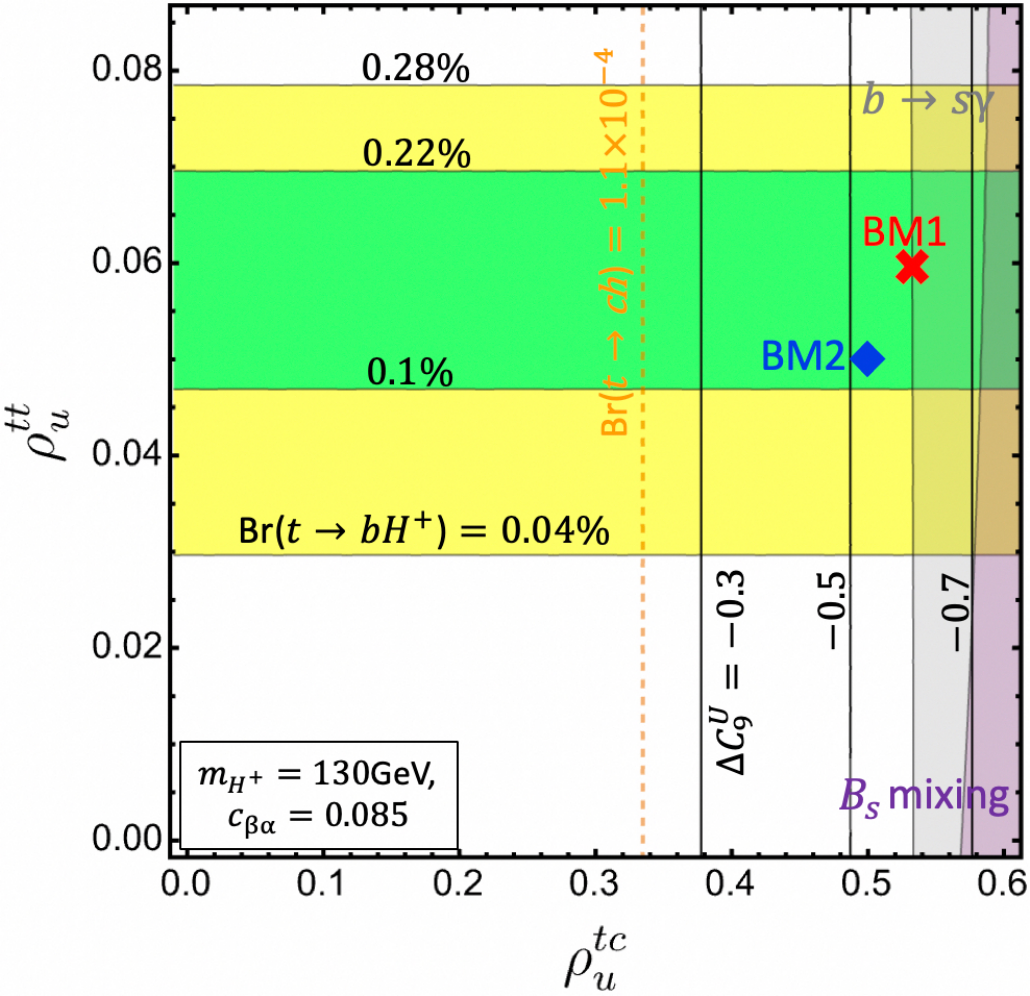}
~~~%
 \includegraphics[width=0.475\textwidth]{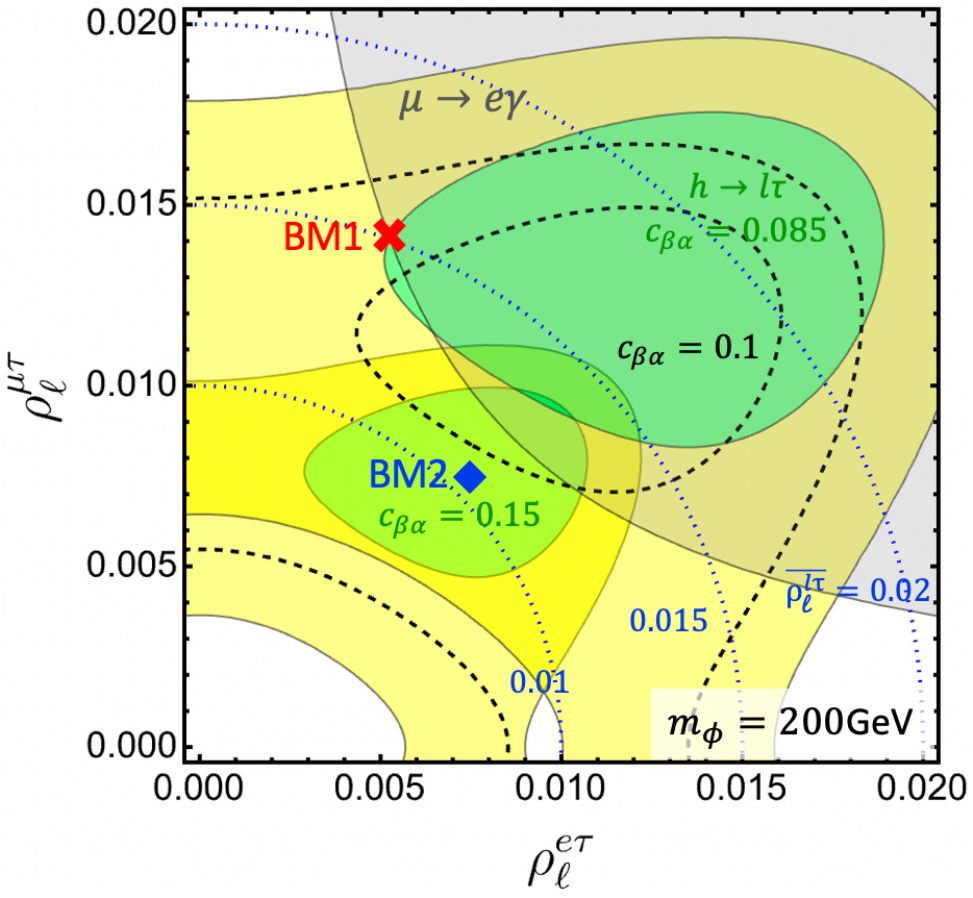}
~~\\
     \vspace{.2cm}
 \includegraphics[width=0.46\textwidth]{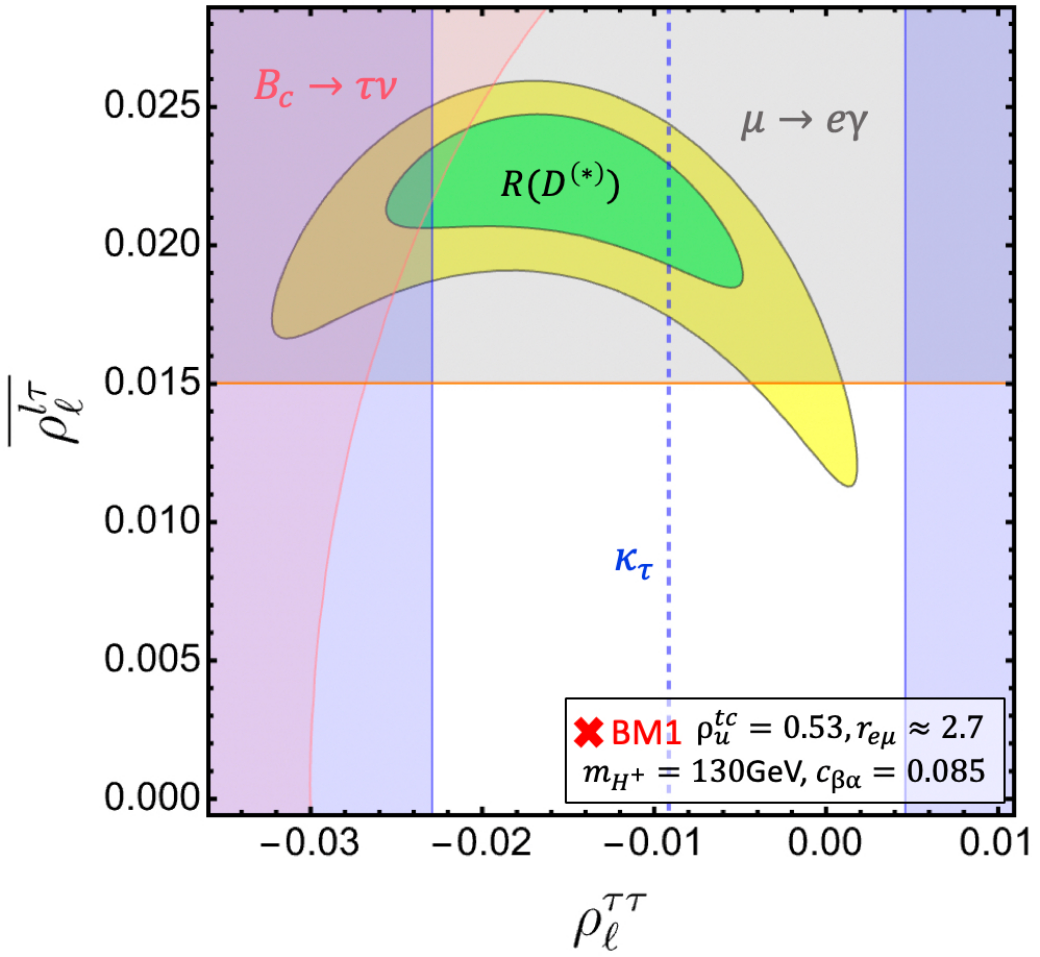}
~~~%
 \includegraphics[width=0.46\textwidth]{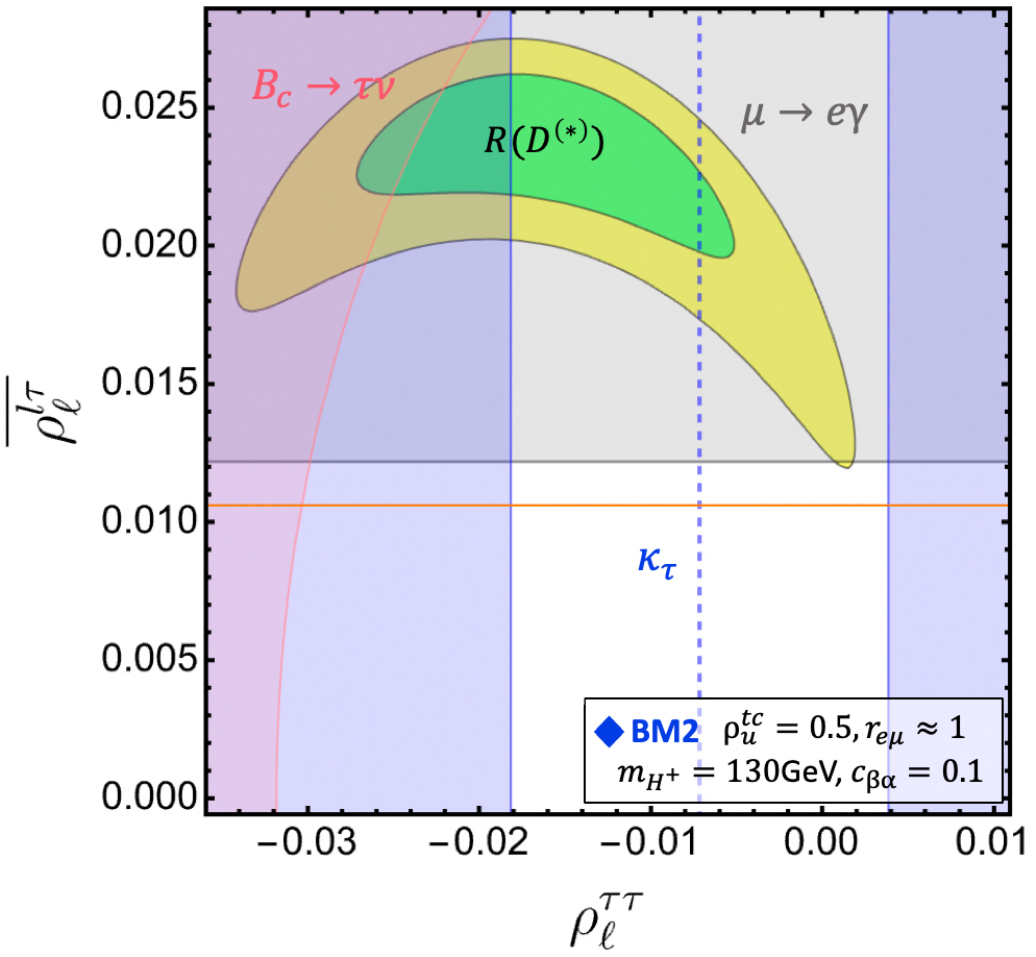}
     \vspace{-.3cm}
\caption{
Upper left: Predicted values of $\Delta C_9^U$ and preferred regions for ${\rm Br}(t\to bH^+)$ in the $\rho_u^{tc}$-$\rho_u^{tt}$ plane along with constraints from $b\to s\gamma$ (lighter gray) and $B_s-\bar B_s$ mixing (darker gray) assuming Br$(H^+\to bc)\approx100\%$. The HL-LHC sensitivity to Br$(t\to hc)$ is shown for $c_{\beta\alpha}=0.085$ (orange-dashed line). The red cross and the blue diamond indicate our two benchmark points BM1 and BM2, respectively.
Upper right: Preferred regions ($1\sigma$ and $2\sigma$) from $h\to l\tau$ for $c_{\beta\alpha}=0.085$, $c_{\beta\alpha}=0.1$ and $c_{\beta\alpha}=0.15$, in the $\rho_\ell^{e\tau}$-$\rho_\ell^{\mu\tau}$ plane as well as the exclusion region from $\mu\to e\gamma$ which, in a linear approximation, is independent of $c_{\beta\alpha}$.
Bottom left (right): Preferred regions from $R({D^{(*)}})$ ($1\sigma$ and $2\sigma$) as well as the exclusion region from $\mu\to e\gamma$ (gray), $\kappa_\tau$ (blue) and $B_c\to\tau\nu$ (red) in the $\rho_\ell^{\tau\tau}$-$\ov{\rho_\ell^{l\tau}}$ plane assuming all Yukawa couplings to be real. The up-quark Yukawa couplings are set to the values of BM1 (BM2) given in the upper figures while the benchmark value of $\ov{\rho_\ell^{l\tau}}$ is indicated by the orange line. The current measured central value of $\kappa_\tau$ is shown as a dashed blue line.}
    \label{fig:yukawa}
    \vspace{-.6cm}
\end{center}
\end{figure*}

\subsection{Collider searches}

Search for the supersymmetric partners of the tau lepton at LHC can constrain $R({D^{(*)}})$ explanations with light charged Higgses~\cite{Iguro:2022uzz}. However, for the coupling hierarchy $\rho_u^{tc}\gg \rho_\ell^{\ell\tau}$, $H^+$ dominantly decays into $c\ov{b}$. In this case, di-jet searches, especially bottom flavoured ones, are relevant. However, the current upper limit on $\rho_u^{tc}$ is a factor 2 weaker than the one favoured by $\Delta C_9^U$~\cite{Iguro:2022uzz,Desai:2022zig} once the $b\to s\gamma$ and $B_s-\bar B_s$ mixing constraints are taken into account. 

Recently, ATLAS and CMS released the result of a same-sign top search targeting a G2HDM~\cite{ATLAS:2023tlp,CMS:2023fod}. They do not find a significant excess in mass range of our interest and set the upper limit, assuming a single scalar particle as a mediator, of $|\rho_u^{tc}|\lesssim 0.3$.\footnote{This constraint could be potentially extrapolated to lower masses. However, this needs detailed experimental analysis since a top quark will be less boosted and this would make the detection more difficult.} However, in the limit of $m_H=m_A$ the effect is suppressed due to destructive interference. 

For $m_\phi\lesssim m_t+m_c$, neutral scalars produced via an EW Drell-Yan process (i.e.~$pp \to Z^*\to AH$) could in principle have sizable decays to $\ell\tau$. This setup would then be stringently constrained by chargino and neutralino searches~\cite{CMS:2021edw,ATLAS:2022nrb,Iguro:2023jkf}. Therefore, these searches can exclude regions in parameter space in which both neutral scalar masses are lighter than $m_t+m_c$ and where Br$(H,A\to l \tau )\sim 1$. However, note that in our setup Br$(H^{+}\to \ov{\tau}\nu)\approx 0$ due to the large $H^{+}\to \ov b c$ width.

\subsection{Higgs coupling strength}

Although both ATLAS~\cite{ATLAS:2022vkf} and CMS~\cite{CMS:2022dwd} found that the $h\tau\ov{\tau}$ coupling is consistent with the SM prediction within uncertainties, their central value is slightly smaller than the SM one, resulting in a coupling strength relative to the SM one of
{\small{
\begin{align}
\kappa_\tau=0.93\pm0.07\,({\rm{ATLAS}}),~~\kappa_\tau=0.92\pm0.08 \,({\rm{CMS}}).
\end{align}
}}
Since nonzero $c_{\beta\alpha}$ and $\rho_\ell^{\tau\tau}$ are necessary for $h\to l\tau$ and $R({D^{(*)}})$, a deviation in $\kappa_\tau$ is inevitable.

In our model, the signal strength $\kappa_\tau$ is given as 
\begin{align}
    \kappa_\tau= \biggl{|}\frac{ \frac{\sqrt{2}m_\tau}{v} s_{\beta\alpha}+\rho_\ell^{\tau\tau} c_{\beta\alpha} }{\frac{\sqrt{2}m_\tau}{v}} \biggl{|}.
\end{align}
Remarkably $c_{\beta\alpha}=0.1$ and $\rho_\ell^{\tau\tau}=-0.01$ give $\kappa_\tau=0.90$ which improves the fit.

\subsection{Oblique correction (ST parameters)}

The mass differences between new Higgses induce deviations of the S and T parameters from 0, i.e.~their SM values. Lead by the CDF-II measurement~\cite{CDF:2022hxs} we have 
\begin{align}
{\rm{S}}=0.086\pm0.077,~~{\rm{T}}=0.177\pm0.070,
\label{eq:2023_ST}
\end{align}
with the correlation of $\rho=0.89$ based on the global fit~\cite{deBlas:2022hdk}.

\section{Phenomenological Analysis}
\label{sec:pheno}

We can now consider the preferred size of the relevant free parameters $\rho_u^{tt},\,\rho_u^{tc},\,\rho_\ell^{\tau\tau},\,\rho_\ell^{\mu\tau}$,\,$\rho_\ell^{e\tau}$, $c_{\beta\alpha}$, $m_{H,A}$ and  $m_{H^+}$, assuming in the first step that all couplings are real and that the other new Yukawa couplings are negligibly small.

Concerning observables that are only sensitive to the charged Higgs contribution, we first use the excess in $t\to H^+b \to b \ov{b}c$ to fix $m_{H^+}\approx 130\,$GeV. Furthermore, $b\to s\ell^+\ell^-$ favors sizable and negative $\Delta C_9^U$ which can be obtained via $\rho_u^{tc}$, such that for Br$(H^+\to c\bar b)\approx 100\%$, i.e.~$|\rho_\ell^{\ell\tau}|,|\rho_u^{cc}|\ll|\rho_u^{tc}|$, leading to $\rho_u^{tt}\approx0.06$. However, the possible effect in $\Delta C_9^U$ is limited to $\approx -0.6$ by the constraints from $b\to s\gamma$ and $B_s-\bar B_s$ mixing. This is illustrated in Fig.~\ref{fig:yukawa} (upper left). Note that the impact of the neutral Higgses can be disregarded for these observables, such that we can choose two benchmark (BM) points for the couplings $\rho_u^{tc}$, i.e.~$\rho_u^{tc}=0.53$ and $\rho_u^{tc}=0.5$ for BM1 and BM2, respectively for $\rho_{u}^{cc}\approx0$.

Turning to observables sensitive to lepton couplings, the excesses in $h\to e \tau $ and $h\to  \mu \tau $ lead to a preference of nonzero values of $\rho_\ell^{e\tau}$, $\rho_\ell^{\mu\tau}$, and $c_{\beta\alpha}$.\footnote{In principle also $\rho_\ell^{\tau e}$, $\rho_\ell^{\tau\mu}$ could explain $h\to e\tau $ and $h\to \mu\tau $. However, in order to avoid chirally enhanced effects in $\mu\to e\gamma$ it is important that both $\rho_\ell^{\tau e} \rho_\ell^{\mu\tau}$ and $\rho_\ell^{e \tau } \rho_\ell^{\tau\mu}$ are not sizable. Furthermore, to avoid effects in $b\to c l\nu$, we will opt for $\rho_\ell^{l \tau}\neq 0$ and $\rho_\ell^{\tau l}=0$.} This at the same time leads to an effect in $\mu\to e\gamma$ as illustrated in Fig.~\ref{fig:yukawa} (upper right). Note the mild dependence on the neutral Higgs masses which we set for definiteness to $200\,$GeV, and that explaining both $h\to e\tau$ and $h\to \mu\tau$ at the same time is possible with a Higgs mixing of $c_{\beta\alpha}\gtrsim 0.08$. Since the significance of excesses in $h\to e\tau$ and $h\to\mu\tau$ are slightly different, the contours are not symmetric in the $\rho_\ell^{e\tau}$-$\rho_\ell^{\mu\tau }$ plane. To maximize the contribution to $R({D^{(*)}})$ while explaining $h\to l\tau$ at $1\sigma$ we fixed the $r_{\mu e}\equiv \rho_\ell^{\mu\tau}/\rho_\ell^{e\tau}=2.7$ and $\ov{\rho_\ell^{l\tau}}\equiv \sqrt{|\rho_\ell^{e\tau}|^2+|\rho_\ell^{\mu\tau}|^2}\approx0.015$ (BM1). In a more conservative setup we use and $r_{\mu e}=1$ and $\ov{\rho_\ell^{l\tau}}=0.011$ (BM2). Finally, since $|\rho_\ell^{l\tau}|\ll|\rho_u^{tc}|$ the results discussed in the previous paragraph are not affected.
   
Let us now consider $R({D^{(*)}})$ in the lower panel in Fig.~\ref{fig:yukawa} for BM1 (left) and BM2 (right) where we also show the $\mu\to e \gamma$ exclusion region in the $\rho_\ell^{\tau\tau}$-$\ov{\rho_\ell^{l\tau}}$ plane. The red and blue regions are excluded by the $B_c\to\tau\nu$ lifetime and $\kappa_\tau$, respectively. Note that the minimal deviation of $\kappa_\tau$ from unity is $4\,\%$ for BM1 since $|\rho_\ell^{\tau\tau}|\gtrsim 5\times 10^{-3}$ and $c_{\beta\alpha}\gtrsim 0.08$ are necessary to explain $h\to l\tau$ and $R({D^{(*)}})$ simultaneously. 

\begin{figure}[t]
    \centering
 \includegraphics[width=0.44\textwidth]{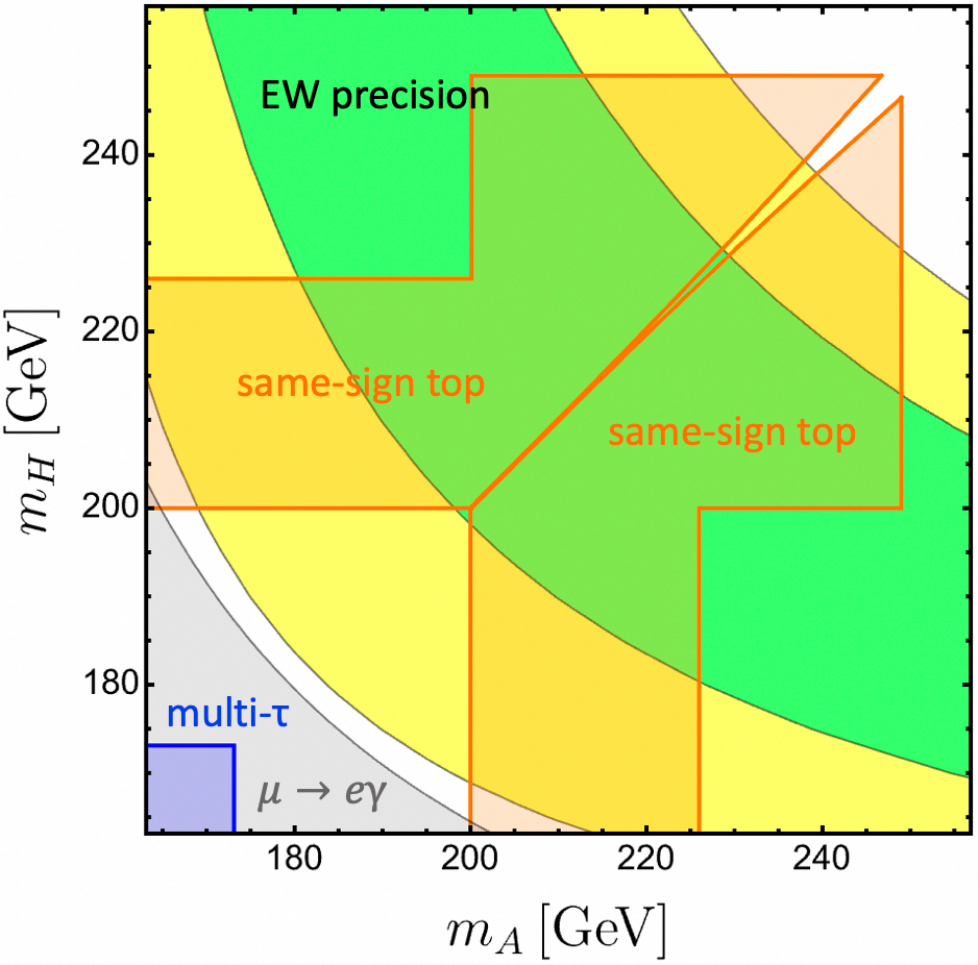}
     \vspace{-.2cm}
\caption{Preferred regions (green:~$1\sigma$, yellow:~$2\sigma$) from electroweak precision data along with exclusion regions from multi-tau and same-sign top searches as well as $\mu\to e \gamma$ in the $m_A$-$m_H$ plane.
}
\label{fig:mAmH}
   \vspace{-.3cm}
\end{figure}

\begin{table}[b]
\begin{center}
\scalebox{1.}{
  \begin{tabular}{c} 
  \hline
    Parameters \\ \hline
    $m_{H^+}=130\,{\rm{GeV}},\,m_\phi=200\,{\rm{GeV}},\,c_{\beta\alpha}=0.1,\,\rho_u^{tt}=0.06$, \\  
    $\rho_u^{tc}=0.47,\,\rho_\ell^{\tau\tau}=-0.01(1\pm 1.8i),\,\rho_\ell^{\mu\tau}=0.01,\,\rho_\ell^{e\tau}=0.006$\\ \hline
   Predictions  \\ \hline
   ${\rm{Br}}(t\to b\ov{b}c)=0.16\%,\,\Delta C_9^U=-0.47,\,{\rm{Br}}(h\to \mu\tau)=0.061\%$,\\
   ${\rm{Br}}(h\to e\tau)=0.022\%,\,R(D)=0.341,\,R(D^*)=0.272,$\\
   $\kappa_\tau=0.91,\,\chi^2_{\rm{SM}}-\chi^2_{\rm{G2HDM}}({\rm{ST}},\,2023)=10.4,$\\
   ${\rm{Br}}(\mu\to e\gamma)=2.0\times 10^{-13},$
   $\Delta C_7=-0.027,\,R_{B_s}=-0.03,$\\
   ${\rm{Br}}(B_c\to\tau\ov\nu)=30\,\%,\,{\rm{Br}}(t\to ch)=3.0\times 10^{-4}$\\ \hline
  \end{tabular}
  }
  \caption{The value of the parameters for BM3 and the corresponding predictions for the observables.
 }
  \label{tab:param_set}
\end{center}   
\vspace{-.25cm}
\end{table}

Note that the BM1 scenario is on the edge of the current constraints such that it can explain all anomalies as well as possible. However, we found that an explanation of $R_{D^{(*)}}$ is possible only within $2\sigma$ level. On the other hand, the BM2 scenario is more conservative w.r.t.~the experimental bounds but is only in agreement with $R({D^{(*)}})$ at the boundary of the $2\sigma$ level. The reason for this is that $\rho_\ell^{e\tau}$ and $\rho_\ell^{\mu\tau}$ are smaller which reduces the non-interfering effect NP with the SM. Since also an imaginary part of $\rho_u^{tc}\rho_\ell^{\tau\tau}$ leads to an amplitude which does not interfere with the SM in $b\to c\tau\nu$, this can help to explain $R({D^{(*)}})$.\footnote{Note that electroweak baryogenesis could be realized with complex Yukawa couplings~\cite{Fuyuto:2017ewj, Kanemura:2023juv}.} We can include the imaginary part of  $\ov{\rho_\ell^{l\tau}}$ as $\rho_\ell^{\tau\tau}$ into the definition of $\ov{\rho_\ell^{l\tau}}$, i.e.~$\ov{\rho_\ell^{l\tau}}=\sqrt{|\rho_\ell^{e\tau}|^2+|\rho_\ell^{\mu\tau}|^2+{\rm{Im}}[\rho_\ell^{\tau\tau}]^2}$.\footnote{Note that $\rho_\ell^{\tau\tau}$ does not contribute to $\mu\to e\gamma$. For simplicity we consider the complex $\rho_\ell^{\tau\tau}$ and assume that $\rho_u^{tc}$ remains to be real. However, $\rho_u^{tc}$ could be complex as well without conflicting $\Delta \Gamma_{B}$.} Once we consider complex $\rho_\ell^{\tau\tau}$, we can generate an imaginary value of the $h\tau\ov\tau$ coupling (for $c_{\beta\alpha}\neq 0$). Since the ATLAS measurement of the SM-Higgs CP properties only starts to constrain this~\cite{ATLAS:2022akr}, the resulting bound is too weak to be relevant. Therefore, $|\rho_\ell^{\tau\tau}|$ can be bigger than in the case $\rho_\ell^{\tau\tau}$ is real and thus explain $R({D^{(*)}})$ with a smaller $|\rho_u^{tc}|$ and hence yield a smaller value of $\Delta C_7$, alleviating the $b\to s\gamma$ bound. The corresponding benchmark point (BM3) is given in Tab.\,\ref{tab:param_set} which explains $t\to bH^+\to b\ov{b}c$, $h\to l\tau$ and $R({D^{(*)}})$ within $1\sigma$ with $\Delta C_9^U\simeq-0.5$ and moderate $\Delta C_7$ and Br$(\mu\to e\gamma)$. 

\begin{figure}[t]
    \centering
 \includegraphics[width=0.44\textwidth]{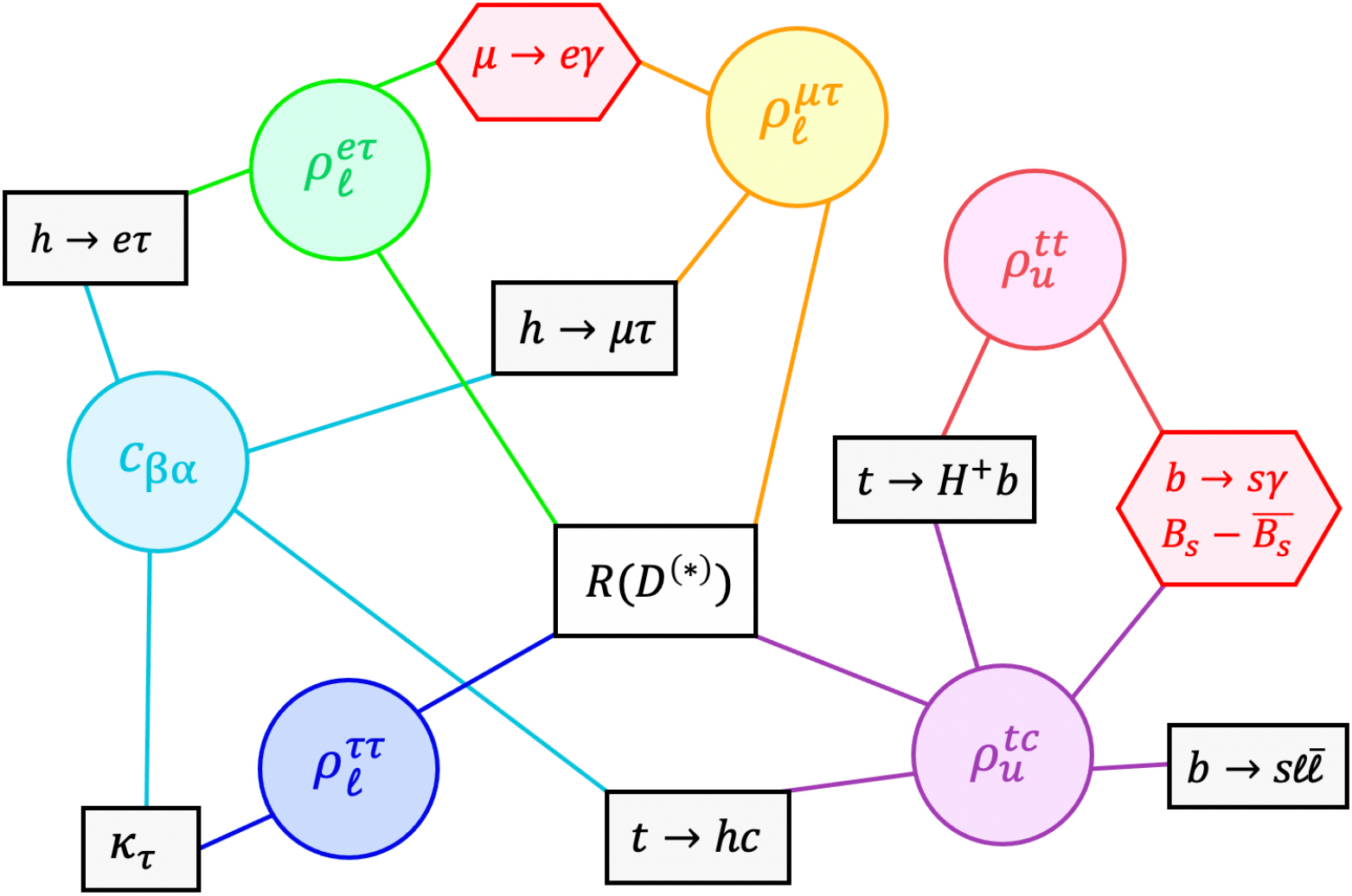}
     \vspace{-.cm}
\caption{Diagram showing the correlations between the free parameters (circles) of our model (except the Higgs masses) and the observables. Observables providing strong constraints are shown as red hexagons while the ones pointing towards a NP effect are shown as black rectangles.}   
\label{fig:correlation}
   \vspace{-.4cm}
\end{figure}

Finally, we consider the impact of varying $m_A$ and $m_H$ in Fig.\,\ref{fig:mAmH} for BM3. Multi-tau final state searches exclude the bottom-left\footnote{Note that the inclusive di-$\tau$ resonance search~\cite{CMS:2022goy} will be able to cover the region where either $H$ or $A$ is lighter than $m_t+m_c$ in future.} part of the $m_A$-$m_H$ plane and small values of $m_A$ and $m_H$ are also disfavoured by Br$(\mu\to e\gamma)$. Same-sign top searches provide constraints if $m_H,\,m_A \gtrsim 200\,$GeV. However, because of the cancellation between the amplitudes from $A$ and $H$, $m_H\simeq m_A$ can evade this bound. Furthermore, once $\phi\to W^\pm H^\mp$ becomes kinematically allowed, same-sign top searches lose their constraining power. Note that top associated Higgs production~\cite{Iguro:2023jju} and bottom associated $H^+$ production~\cite{Blanke:2022pjy} as well as lowering the threshold of same-sign top searches~\cite{ATLAS:2023tlp,CMS:2023fod} are crucial to probe this scenario.

\section{Conclusions and discussion}
\label{sec:conclusion}

Motivated by the hints for NP in $t\to bH^+$, $b\to s\ell^+\ell^-$, $h\to e\tau$, $h\to\mu\tau$, $m_W$ and $R({D^{(*)}})$ we revisited the model with the minimal particle context that is potentially capable of providing a combined explanation, the 2HDM with generic sources of flavour violation. Even though the model is very predictive and hence constrained, we found a minimal set of parameters (Fig.\,\ref{fig:correlation}) that can address these deviations from the SM predictions simultaneously without violating any other bounds. For this, a mild mass difference between the charged and additional neutral Higgs is necessary to evade the LHC constraint, at the same time improving the EW global fit by shifting the prediction for the $W$ mass. Furthermore, a deviation in the SM Higgs coupling strength to tau leptons $\kappa_\tau$ and a non-zero rate for $t\to hc$ are predicted, both welcomed by current data. 

While we assumed the other Yukawa coupling to be negligible, $\rho_d^{bb}\approx\mathcal{O}(10^{-2})$ could be helpful to reduce the effect in $\Delta C_7$ while allowing for $b$-associated production of the new neutral scalars at the LHC. Adding a small $\rho_u^{cc}$ would induce $\Delta C_9^U$ (see Eq.\,\ref{eq:C9U}).\footnote{It would be important to comment that an additional $\rho_u^{cc}$ does not induce $D-\bar {D}$ mixing since $H^{+}$ does not couple to up quark in our setup.} Note that once we give up either $h\to\mu\tau$ or $h\to e\tau$, the $\mu \to e \gamma$ constraint can be relaxed such that $R({D^{(*)}})$ could be fully explained. This is because $\rho_\ell^{e\tau}$ or $\rho_\ell^{\mu\tau}$ can be larger and hence the smaller $c_{\beta\alpha}$ is allowed. Then larger $\rho_\ell^{\tau\tau}$ and smaller $\rho_u^{tc}$ can explain $R({D^{(*)}})$. While a smaller $\rho_u^{tc}$ would lead to a smaller contribution to $\Delta C_9^U$, a tiny $\rho_u^{cc}$ can already regenerate a sizable value. Note that a smaller $\rho_u^{tc}$ would also be beneficial to avoid tuning the neutral Higgs masses while still avoiding collider constraints. To assess the validity of such a more complicated scenario, a global fit, e.g.~with the public tool GAMBIT~\cite{Athron:2021auq}, is desirable for future research.

\begin{acknowledgments} 
We are very grateful to Lisong Chen, Marco Fedele, Ulrich Nierste, Teppei Kitahara, Hiroyasu Yonaha and Martin Lang for enlightening discussions and encouraging this work. The work of A.\,C.~is supported by a professorship grant from the Swiss National Science Foundation (No.\ PP00P21\_76884). S.\,I. would like to thank PSI for the warm hospitality where he stayed during the initial stage of this project. S.\,I. is supported by the Deutsche Forschungsgemeinschaft (DFG, German Research Foundation) under grant 396021762-TRR\,257. 
\end{acknowledgments}

\bibliographystyle{utphys}
\bibliography{references}
\end{document}